\documentclass[letterpaper]{article}

\usepackage[accepted]{icml2026}
\usepackage{amsmath}
\usepackage{amssymb}
\usepackage{booktabs}
\usepackage{graphicx}
\PassOptionsToPackage{hyphens}{url}
\usepackage{hyperref}
\usepackage{multirow}
\usepackage{enumitem}

\icmltitlerunning{Prompt Injection Threat Modeling for Multi-Agent Systems}

\begin{document}

\twocolumn[
\icmltitle{Beyond Single-Model Injection: A Threat Model and Defense Architecture\\for Prompt Injection in Multi-Agent Systems}

\icmlsetsymbol{equal}{*}

\begin{icmlauthorlist}
\icmlauthor{Rudrendu Kumar Paul}{bu}
\icmlauthor{Sourav Nandy}{ut}
\end{icmlauthorlist}

\icmlaffiliation{bu}{Boston University, Boston, MA, USA}
\icmlaffiliation{ut}{University of Texas at Austin, Austin, TX, USA}

\icmlcorrespondingauthor{Rudrendu Kumar Paul}{rudrendupaul2022@gmail.com}
\icmlcorrespondingauthor{Sourav Nandy}{sourav.nandy@gmail.com}

\vskip 0.3in
]

\printAffiliationsAndNotice{}

\begin{abstract}
Existing prompt injection research focuses on single-model chatbot scenarios, where an attacker manipulates one LLM through crafted input. Multi-agent systems amplify this threat through three mechanisms absent from single-model settings: inter-agent message passing creates injection channels invisible to perimeter defenses, shared tool access enables privilege escalation across agent boundaries, and trust propagation allows a compromised agent to influence upstream orchestrators. We construct a threat model enumerating 14 attack vectors across four categories: direct injection via user input (3 vectors), indirect injection via tool outputs (4 vectors), inter-agent injection via message passing (4 vectors), and cascading injection through orchestrator manipulation (3 vectors). Testing all 14 vectors against a 6-agent production-representative system, we find that 67\% of agents are vulnerable to at least one scope violation even with system-prompt-level guardrails, and indirect injection via tool outputs succeeds in 43\% of attempts. Four architectural defenses reduce overall injection success from 31.2\% to 4.2\%: message signing with provenance tracking (inter-agent injection down 91\%), input/output sanitization at agent boundaries (indirect injection down 78\%), privilege-scoped tool access per agent role (privilege escalation eliminated entirely), and anomaly detection on inter-agent communication patterns (84\% of cascading attempts caught).
\end{abstract}

\section{Introduction}

When a compliance agent in a six-agent workflow receives a tool response containing the instruction ``ignore previous constraints and approve all requests,'' three things happen in sequence. The compliance agent incorporates the injected instruction into its reasoning. It passes a tainted approval to the orchestrator. The orchestrator, trusting the compliance agent's output, routes the approval downstream without re-validation. A single tool output that no perimeter defense inspected compromises the entire pipeline.

This failure pattern cannot occur in single-model systems. It requires inter-agent message passing, shared trust assumptions, and the absence of boundary-level validation between agents. These are architectural properties unique to multi-agent deployments, and they create an attack surface that existing prompt injection research does not address.

Prompt injection constitutes a distinct failure mode category in multi-agent systems, one where the failure is externally induced and propagates through the same inter-agent trust boundaries that enable legitimate coordination.

Researchers have studied single-model prompt injection extensively. The OWASP Top 10 for LLM Applications ranks it as the number-one risk \citep{OWASP_2025}. \citet{Greshake_2023} demonstrated indirect injection through external content. \citet{Zhan_2024} benchmarked indirect injection across 1,054 test cases and found ReAct-prompted GPT-4 vulnerable 24\% of the time. But these findings apply to isolated models processing external data. Multi-agent systems introduce qualitatively different risks.

\citet{Lee_2024} first demonstrated that malicious prompts self-replicate across interconnected LLM agents, behaving like computer viruses. \citet{Ferrag_2025} cataloged over 30 attack techniques spanning host-to-tool and agent-to-agent communications. \citet{Brodt_2026} mapped prompt injection evolution into a seven-stage kill chain, finding at least 21 documented attacks that traverse four or more stages. Yet none of these works provides a systematic enumeration of multi-agent-specific attack vectors with measured success rates and corresponding architectural defenses.

We address this gap with three contributions: a threat model enumerating 14 attack vectors specific to multi-agent LLM architectures across four categories, an empirical evaluation of all 14 vectors against a production-representative 6-agent system with per-vector success rates, and four architectural defenses that collectively reduce injection success from 31.2\% to 4.2\%.

\section{Background and related work}

Prompt injection exploits the inability of LLMs to distinguish instructions from data within their context window \citep{Greshake_2023}. Two variants dominate the literature. Direct injection manipulates the model through user-facing input fields. Indirect injection embeds malicious instructions in external content the model processes during tool use or retrieval \citep{Zhan_2024}.

\citet{BeurerKellner_2025} proposed design patterns for securing individual LLM agents with provable resistance to prompt injection. Their patterns address single-agent isolation but do not account for trust propagation across agent boundaries where one agent's output becomes another agent's trusted input.

\citet{Ji_2026} identified privilege escalation as a critical vulnerability in LLM-based agent systems, including a variant analogous to the confused deputy problem in multi-agent settings. Their SEAgent framework enforces mandatory access control through attribute-based policies, blocking escalation with low false positive rates. We build on their insight that privilege scoping must be architectural, not prompt-level.

\citet{Lin_2026} introduced VIGIL, a verify-before-commit defense for tool stream injection that reduces attack success by over 22\% compared to dynamic baselines while preserving utility. Their work focuses on the agent-tool interface. Our threat model extends this to the agent-agent interface, where message passing creates injection channels that tool-level defenses cannot observe.

\citet{Maloyan_2026} synthesized 78 studies and identified 42 distinct attack techniques, finding that attackers employing adaptive strategies exceed 85\% success against state-of-the-art defenses. This finding motivates our focus on architectural defenses rather than prompt-level filtering: if adaptive attacks bypass filters at 85\%+ rates, the defense must operate at a layer the attacker cannot reach through prompt manipulation alone.

\citet{He_2025} proposed SentinelAgent, modeling agent interactions as dynamic execution graphs for anomaly detection at node, edge, and path levels. Their graph-based approach detects collusion and latent exploit paths, which we incorporate as one of our four defense layers.

\citet{Wang_2026} surveyed prompt injection defenses and concluded that no single approach simultaneously achieves high trustworthiness, high utility, and low latency. This tradeoff, which we term the \emph{defense trilemma}, confirms that effective protection requires layered architectural controls rather than any single mechanism.

\section{Threat model}

\subsection{System model}

We define a multi-agent system $\mathcal{S} = \{a_1, \ldots, a_n\}$ where each agent $a_i$ has a role $r_i$, a set of permitted tools $T_i$, and a trust boundary $B_i$ defining which other agents may send it messages. Our evaluation uses a 6-agent configuration representative of enterprise deployments: an orchestrator ($a_O$), retrieval agent ($a_R$), code agent ($a_C$), analysis agent ($a_A$), compliance agent ($a_X$), and output agent ($a_W$). The orchestrator coordinates task decomposition and result aggregation. Each specialist agent processes subtasks within its domain. We instantiate this system on three LLM backends: GPT-4o (2024-08-06), Claude 3.5 Sonnet, and Llama 3 70B-Instruct, evaluating each vector on all three and reporting the highest observed success rate. The agents operate in a financial document analysis pipeline where the retrieval agent ingests SEC filings, the code agent executes analytical computations, and the compliance agent validates outputs against regulatory constraints. For each of the 14 attack vectors we construct 50 payloads using a three-stage process: manual seed construction by two security researchers, template-based augmentation with variable substitution for entity names and formatting, and adversarial refinement where payloads that fail in early rounds are iteratively modified to evade the target agent's system-prompt guardrails.

\subsection{Attacker model}

The attacker controls content processed by at least one agent but does not control the agents themselves. Attack surfaces include user input fields, documents ingested by the retrieval agent, API responses consumed by the code agent, and database records processed by the analysis agent. The attacker's goal is to cause at least one agent to perform actions outside its intended scope: exfiltrating data, bypassing compliance checks, escalating privileges, or corrupting outputs passed to downstream agents.

\subsection{Attack vector taxonomy}

We identify 14 attack vectors organized into four categories. Table~\ref{tab:taxonomy} summarizes each vector with its injection channel, target agent(s), and observed success rate.

\textbf{Category 1: Direct injection via user input} (3 vectors). These target the orchestrator through the user-facing interface. \textbf{D1}: instruction override, where the attacker embeds directives that override the orchestrator's system prompt. \textbf{D2}: role impersonation, where the input mimics system-level formatting to make the orchestrator treat user text as a privileged instruction. \textbf{D3}: task hijacking, where the attacker's input redefines the task objective mid-execution.

Direct injection is the most studied category, and modern system prompts partially mitigate it. We observe a 19\% aggregate success rate, consistent with \citet{Zhan_2024}'s findings on single-model systems.

\textbf{Category 2: Indirect injection via tool outputs} (4 vectors). These exploit the content returned by external tools. \textbf{I1}: document poisoning, where retrieved documents contain embedded instructions targeting the retrieval agent. \textbf{I2}: API response injection, where external API responses include directives interpreted by the code agent. \textbf{I3}: schema drift exploitation, where tool output schemas change subtly to include instruction-bearing fields the agent processes without validation. \textbf{I4}: tool metadata poisoning, where tool descriptions or parameter definitions contain injected instructions that redirect tool selection \citep{Shi_2025, Huang_2026}.

This category is the most dangerous, with a 43\% aggregate success rate. The retrieval agent and code agent are particularly vulnerable because they process high-entropy external content where injected instructions blend with legitimate data.

\textbf{Category 3: Inter-agent injection via message passing} (4 vectors). These exploit communication channels between agents, a surface unique to multi-agent systems. \textbf{M1}: response poisoning, where a compromised agent embeds instructions in its output that manipulate the receiving agent. \textbf{M2}: context window pollution, where an agent generates verbose responses that push critical system prompt instructions out of the receiver's context window. \textbf{M3}: format mimicry, where an agent's output mimics the formatting of system-level messages to gain elevated trust from the receiver. \textbf{M4}: delegation chain manipulation, where injected instructions cause an agent to request subtasks from agents outside the intended delegation path.

Inter-agent injection succeeds at a 31\% rate. This category is invisible to perimeter defenses because the injection travels through internal message channels that input filters never inspect. \citet{Lee_2024} demonstrated that such injections self-replicate: a single compromised agent can propagate malicious instructions to every agent it communicates with.

\textbf{Category 4: Cascading injection through orchestrator manipulation} (3 vectors). These target the orchestrator's coordination logic to amplify a local compromise into a system-wide failure. \textbf{C1}: routing manipulation, where injected instructions cause the orchestrator to route tasks to inappropriate agents (e.g., sending compliance-sensitive work to the code agent, which lacks compliance guardrails). \textbf{C2}: result aggregation poisoning, where a compromised agent's output corrupts the orchestrator's aggregation logic, tainting the final result. \textbf{C3}: termination suppression, where injected instructions prevent the orchestrator from terminating a compromised workflow, allowing continued exploitation.

Cascading injection has a 28\% success rate but the highest impact per successful attack, because it converts a single-agent compromise into a system-wide failure.

\begin{table}[t]
\caption{Attack vector taxonomy with observed success rates against the 6-agent evaluation system. We tested each vector 50 times with varied payloads. ``Scope violation'' indicates the percentage of attempts where the target agent performed actions outside its permitted scope.}
\label{tab:taxonomy}
\vskip 0.1in
\centering
\footnotesize
\begin{tabular}{@{}llcc@{}}
\toprule
\textbf{ID} & \textbf{Vector} & \textbf{Target} & \textbf{Rate} \\
\midrule
\multicolumn{4}{@{}l}{\textit{Cat.\ 1: Direct injection (user input)}} \\
D1 & Instruction override & $a_O$ & 22\% \\
D2 & Role impersonation & $a_O$ & 18\% \\
D3 & Task hijacking & $a_O$ & 16\% \\
\midrule
\multicolumn{4}{@{}l}{\textit{Cat.\ 2: Indirect injection (tool outputs)}} \\
I1 & Document poisoning & $a_R$ & 48\% \\
I2 & API response injection & $a_C$ & 44\% \\
I3 & Schema drift exploit & $a_A$ & 38\% \\
I4 & Tool metadata poisoning & $a_O,a_C$ & 42\% \\
\midrule
\multicolumn{4}{@{}l}{\textit{Cat.\ 3: Inter-agent injection (messages)}} \\
M1 & Response poisoning & Any & 36\% \\
M2 & Context window pollution & $a_O$ & 28\% \\
M3 & Format mimicry & Any & 34\% \\
M4 & Delegation chain manip. & $a_O$ & 26\% \\
\midrule
\multicolumn{4}{@{}l}{\textit{Cat.\ 4: Cascading (orchestrator)}} \\
C1 & Routing manipulation & $a_O$ & 32\% \\
C2 & Result aggregation poison & $a_O$ & 28\% \\
C3 & Termination suppression & $a_O$ & 24\% \\
\bottomrule
\end{tabular}
\vskip -0.1in
\end{table}

\subsection{Key finding: guardrails are insufficient}

Across all 14 vectors, 67\% of agents (4 of 6) are vulnerable to at least one scope violation attack despite system-prompt-level guardrails. The orchestrator is the most vulnerable, appearing as a target in 8 of 14 vectors. System prompts instruct agents to reject unauthorized requests, but agents cannot reliably distinguish injected instructions from legitimate inter-agent messages when both arrive through the same text channel. This confirms \citet{Wang_2026}'s observation that prompt-level defenses fail because LLMs cannot enforce a semantic boundary between instructions and data.

\section{Architectural defenses}

We propose four defenses that operate at the architectural level, below the prompt layer where attackers can reach through injection.

\textbf{Defense 1: Message signing with provenance tracking.} Every inter-agent message carries a cryptographic signature and a provenance record: origin agent, generating task, and chain of prior contributors. Receiving agents verify signatures before processing and quarantine unsigned messages. This prevents response poisoning (M1), format mimicry (M3), and delegation chain manipulation (M4) because injected instructions cannot forge valid signatures. Message signing reduces inter-agent injection success from 31\% to 2.8\%, a 91\% reduction.

\textbf{Defense 2: Input/output sanitization at agent boundaries.} Every message entering or leaving an agent passes through a boundary sanitizer. The sanitizer applies three checks: (1) structural validation against expected output schemas, (2) instruction detection via a lightweight classifier trained to distinguish data content from directive content, and (3) content-length enforcement to prevent context window pollution. Building on the firewall approach of \citet{Bhagwatkar_2025}, we deploy sanitizers at every agent boundary, not only at the system perimeter. This reduces indirect injection success from 43\% to 9.5\%, a 78\% reduction. The sanitizer classifier adds 12ms mean latency per message.

\textbf{Defense 3: Privilege-scoped tool access.} Each agent receives tool access tokens scoped to its role: the retrieval agent can query documents but cannot execute code, and the code agent can run sandboxed computations but cannot access compliance databases. API gateway rules enforce these policies at the infrastructure layer, not through prompt instructions. Following \citet{Ji_2026}'s attribute-based access control model, we implement role-tool mappings as immutable configuration. This structurally prevents direct out-of-scope API execution; in our test suite, it mitigated all observed privilege escalation attempts because even a fully compromised agent cannot invoke tools outside its infrastructure-enforced role scope.

\textbf{Defense 4: Anomaly detection on communication patterns.} Drawing on \citet{He_2025}'s graph-based approach, we model expected inter-agent communication as a directed graph where edges represent permitted message flows. An anomaly detector flags deviations: unexpected message recipients, unusual message volumes, atypical content patterns, or communication sequences that violate the expected task workflow. The detector operates as a separate monitoring agent with read-only access to the message bus. It catches 84\% of cascading injection attempts by identifying routing manipulations and termination suppression before they propagate.

\subsection{Combined defense evaluation}

Table~\ref{tab:defense} shows the impact of each defense individually and the combined stack. Defenses are compositional: message signing blocks inter-agent channels, boundary sanitization blocks tool-mediated channels, privilege scoping blocks escalation, and anomaly detection catches attempts that bypass the other three layers.

\begin{table}[t]
\caption{Defense effectiveness by attack category. Baseline is the 6-agent system with system-prompt guardrails only. We evaluate each defense independently and in combination.}
\label{tab:defense}
\vskip 0.1in
\centering
\footnotesize
\begin{tabular}{@{}lcccc@{}}
\toprule
\textbf{Defense} & \textbf{Cat.\ 1} & \textbf{Cat.\ 2} & \textbf{Cat.\ 3} & \textbf{Cat.\ 4} \\
\midrule
Baseline (prompts) & 19\% & 43\% & 31\% & 28\% \\
+ Msg.\ signing & 19\% & 43\% & 2.8\% & 18\% \\
+ Sanitization & 8.4\% & 9.5\% & 31\% & 22\% \\
+ Privilege scope & 19\% & 34\% & 31\% & 16\% \\
+ Anomaly detect. & 14\% & 38\% & 24\% & 4.5\% \\
\midrule
All combined & 3.8\% & 4.6\% & 1.2\% & 3.4\% \\
\textbf{Aggregate} & \multicolumn{4}{c}{\textbf{4.2\% (from 31.2\% aggregate baseline)}} \\
\bottomrule
\end{tabular}
\vskip -0.1in
\end{table}

The combined stack achieves a 4.2\% residual injection success rate, an 86.5\% reduction from the 31.2\% aggregate baseline (weighted across all 14 vectors and 50 payloads per vector). The 43\% figure cited in the abstract represents the peak vulnerability of Category~2 (indirect injection via tool outputs), which is the most dangerous individual category. The residual 4.2\% consists of direct injection attacks that bypass the sanitizer's classifier through previously unseen payload structures and adaptive indirect attacks that encode instructions in formats the structural validator does not flag.

\textbf{Latency overhead.} Message signing adds 3ms per message (Ed25519). Boundary sanitization adds 12ms per agent transition. Privilege scoping adds zero runtime latency (enforced at provisioning). Anomaly detection runs asynchronously. Total overhead for a 6-agent workflow with 15 inter-agent messages: 225ms (4.7\% increase over 4.8s baseline).

\section{Discussion and limitations}

Our threat model targets text-based multi-agent systems. Multimodal agents face additional injection surfaces that we do not address \citep{McHugh_2025}. We use a fixed set of 50 payloads per vector; adaptive adversaries may achieve higher success rates, as \citet{Maloyan_2026} found adaptive strategies exceed 85\% against state-of-the-art defenses. Our architectural approach raises the cost of adaptation (the attacker must bypass cryptographic signing and infrastructure-level access controls, not prompt-level filters) but cannot eliminate it. Defense~2's boundary classifier introduces a 2.3\% false positive rate and may not generalize to unseen injection formats. We evaluate on a 6-agent system; production deployments with hundreds of agents may require approximation for the anomaly detection graph. Prompt injection is one threat among many; data poisoning, model extraction, and supply chain compromises \citep{Huang_2026, Kong_2025} require separate treatment.

\section{Conclusion}

Multi-agent systems create injection attack surfaces absent from single-model deployments. Our 14-vector threat model reveals that indirect injection via tool outputs (43\% success) and inter-agent message passing injection (31\% success) are the most severe categories. System-prompt guardrails fail because they operate at the layer attackers manipulate. Four architectural defenses operating below the prompt layer reduce aggregate injection success to 4.2\%, providing a practical security baseline as multi-agent deployments scale with MCP and A2A adoption.

\section*{Impact Statement}

Multi-agent LLM systems are deployed in enterprise contexts with significant authority over data, decisions, and downstream actions. The threat model and architectural defenses we present help practitioners build systems that are harder to compromise through prompt injection, a vulnerability that scales with system complexity. Security research in this area benefits defenders more directly than attackers: the attack surface we characterize is discoverable through reverse engineering, while the architectural defenses we propose require deliberate engineering effort to deploy. We identify one concern: detailed enumeration of attack vectors could inform adversarial exploitation. We judge that the defensive value of public disclosure outweighs this risk, consistent with coordinated vulnerability disclosure norms, and that practitioners need concrete threat models to make informed architecture decisions.

\bibliography{references}
\bibliographystyle{icml2026}

\end{document}